\documentclass[aps,prb,floatfix,twocolumn,showpacs]{revtex4}
\usepackage{graphicx}

\usepackage[normalem]{ulem} 
\usepackage{color} 

\newcommand{\ket}[1]{|#1 \rangle}
\newcommand{\bra}[1]{\langle #1|}

\begin{document}
\title{Interferometry using spatial adiabatic passage in
  quantum dot networks}

\author{Lenneke M. Jong}
\email{lmjong@unimelb.edu.au}
\affiliation{Centre for Quantum Computer Technology, School of
  Physics, University of Melbourne, VIC 3010}

\author{Andrew D. Greentree}
\affiliation{School of Physics, University of Melbourne, VIC 3010}

\begin{abstract}
We show that techniques of spatial adiabatic passage can be used to
realise an electron interferometer in a geometry analogous to a
conventional Aharonov-Bohm ring, with transport of the particle
through the device modulated using coherent transport adiabatic passage.  
This device shows an interesting interplay between the adiabatic and
non-adiabatic behaviour of the system.
 The transition between non-adiabatic and adiabatic behaviour may be tuned via system parameters
and the total time over which the protocol is enacted. Interference effects in the final state populations analogous to the electrostatic Aharonov-Bohm effect are observed.
\end{abstract}


\maketitle

\section{Introduction}

Quantum information science offers a wealth of new and potentially
important phenomena to explore. One hope is that these possibilities
will translate into practical and unique devices, with functionalities
unmatched by classical devices.  One field that has explored quantum
mechanical effects and quantum coherence in particular is the field of
light-matter interactions, and especially the couplings and effects
investigated in the physics of optically driven multi-level atoms \cite{VitanovARPC2001}.  Moving from atomic systems, with particular level structures, to spatial systems, perhaps defined lithographically, provides dramatic opportunities to tailor the Hilbert space and connectivity of the resultant quantum system.  This allows natural translations of known effects from the atomic realm, to the electronic realm.  Examples include charge qubits \cite{HollenbergPRB2004} and qudits \cite{GreentreePRL2004}, adiabatic passage \cite{GreentreePRB2004}, and electromagnetically-induced transparency \cite{EmaryPRB2007}.  A review of some of the progress in mesoscopic electronic systems, focussing on the quantum electronics/quantum optics/quantum information interface can be found in Ref.\cite{BrandesPhysRep2005}.

The solid-state provides exceptional possibilities for realizing
spatially defined Hilbert spaces, especially in the context of
superconductors, donors in silicon \cite{ClarkPhilTran2003} and
quantum dots.  Here we focus on the latter for conceptual simplicity,
but our ideas are equally relevant to all these implementations.
Quantum dots provide a rich platform for exploring novel experiments
in quantum mechanics and often allow for fine control over many system
parameters. Chains of connected dots are of particular interest, with
recent studies including examinations of triple quantum dots in GaAs 2DEG structures \cite{GaudreauPRL2006,RoggePRB2008,AmahaAPL2009,ShroerPRB2007,GaudreauPRB2009}, carbon nanotubes \cite{GroveNanoLett2008} and double quantum dots in silicon \cite{LimAPL2009}, which extend the accessible spatial Hilbert space to allow clear connections with the quantum optical case. 

Coherent Tunnelling Adiabatic Passage (CTAP) \cite{GreentreePRB2004} has been proposed for transporting quantum information and is a spatial analogue of the well known STIRAP protocol from quantum optics \cite{VitanovARPC2001}. It transports a particle coherently using a counter-intuitive coupling sequence of tunnelling matrix elements. CTAP has been recently demonstrated by Longhi \textit{et al.} using photons in three and multi-waveguides structure \cite{LonghiJPhysB2007,LonghiPRB2007,DellaValleAPL2008} but it has also been proposed for observation in quantum dots \cite{PetrosyanOptComm2006,ColePRB2008,OpatrnyPRA2009}. Other platforms for which demonstrations of CTAP have been proposed include phosphorus donors in silicon \cite{HollenbergPRB2006, RahmanPRB2009,VanDonkelaar2008}, where CTAP can be used as the transport mechanism in a quantum computer architecture, and also in superconductors \cite{SiewertOptComm2006}, single atoms in optical potentials
\cite{EckertPRA2004, EckertOptCom2006}, and Bose-Einstein
Condensates \cite{GraefePRA2006,RabPRA2008}. STIRAP has also been
discussed for spatial particle motion in quantum dots, see for example
Ref. \cite{RenzoniPRB2001, SiewertAdvSSP2004}.

The extensible nature of controlling the spatial location of states
and their connectivity has led to CTAP being extended  to multiple
recipients (MRAP) \cite{GreentreePRA2006,DevittQIP2007} as a means of
implementing a form of fanout for a quantum computer. This branching
ability, and the fact that CTAP is a coherent process, lends itself to
investigating  interferometry with devices using the CTAP as the
transport mechanism.

Interferometry is a well-known means of probing the wave-like
  properties of particles, and non-trivial quantum phases. There have
  also been solid-state experiments that utilise interferometry to
  observe the wave-like nature of electrons in Aharonov-Bohm (AB) and
  related interferometers for 'which-path' measurements \cite{BuksNature1998,JiNature2003,ChangNature2008}.  The AB effect is a topological effect
  where a charged particle (for example an electron) traverses a loop.
  If the loop encloses a magnetic flux, this will break the symmetry
  of the paths and introduce a non-trivial phase.  Interferometry then
  allows this phase to be revealed in oscillations in the final state
  population, as a function of the enclosed flux.

Here we are considering an adiabatic realization of the
electrostatic Aharonov-Bohm (EAB) effect \cite{AharonovBohmPhysRev1959,BoyerPRD1973}. This effect is
named by analogy with the AB effect, but is not a topological
effect.  Instead, an external electrostatic field is used to break
the symmetry of the two paths, which again gives rise to non-trivial
phases and population oscillations as a function of the field.  The
EAB effect has  been investigated in the context of one dimensional
mesoscopic rings in metallic \cite{WashburnPRL1987} and semiconductor structures \cite{DattaAPL1986,deVegvarPRB1989} and generalised for the case where
both electrostatic and magnetostatic potentials exist
\cite{TakaiPRB1993}. 

\begin{figure}[tb]
\includegraphics[width=0.45\textwidth]{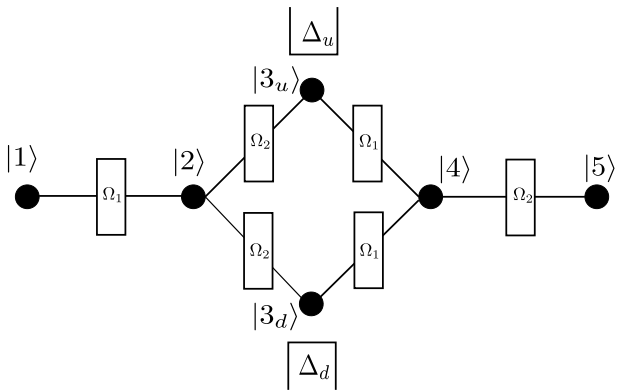}
\caption{The interferometer structure consisting of six coupled
  sites with gate-controlled TMEs $\Omega_1(t)$ and $\Omega_2(t)$. The particle is moved from $\ket{1}$ to $\ket{5}$ 
  using the alternating CTAP protocol via the two intermediate states $\ket{3_u}$ and $\ket{3_d}$.  The control gates, labelled $\Delta_u$ and $\Delta_d$, break the symmetry of the paths, leading to interference effects seen in the final state population as a function of the magnitude of the energy shifts and the total time of the transport. Note that we assume that the energy that the electron has at each site has been set equal. 
\label{fig:ctapint}}
\end{figure}

We show that CTAP can be used to explore physics similar to those
seen in the EAB effect. 
Our suggested geometry, highly reminiscent of traditional AB rings, is shown in Fig.~\ref{fig:ctapint}. It consists of six sites with the tunnelling between
sites controlled via surface control gates. The energies of the two middle sites in each branch may also be controlled  by a gate and it is assumed that the energy of an electron at each other site is also controlled, static, and equal. 
A square quantum dot, such as that examined in Refs. \cite{CreffieldPRB1999,JeffersonPRA2002}, with added coupling to a source and drain, is one possible implementation of this structure. Aharonov-Bohm-like oscillations in an adiabatic
  passage four dot ring in the presence of a magnetic field has also been considered \cite{AlcobiMasters2007}.  

The device shown in Fig.~\ref{fig:ctapint} is analogous to a Mach-Zehnder
interferometer where the final state population observed is dependent
on the interference condition created by the gate-controlled energy shifts on either
arm. Because of the deep analogy between this structure and that of quantum optical systems, we shall refer to the energy shifts of these donors as detunings. The detuning on the middle sites is controlled individually by
gates and plays a similar role to the phase difference between the two arms of the Mach-Zehnder interferometer. Here we are investigating the passage of a single electron travelling through the device in some superposition of the two pathways. Where the detuning is different in the two paths there will be different phase accumulated by the electron, which is exhibited in the interference pattern in the final state population of the end-of-chain site. The device we consider uses adiabatic-passage as the means of transport.  This provides interesting flexibility and a rich phase space to explore. In particular this six-dot geometry is the simplest topology that permits both adiabatic passage and a non-trivial loop.  Furthermore, our model shows an interesting interplay between adiabatic and nonadiabatic features, as is discussed below.

In our proposed quantum dot geometry, the Tunnelling Matrix Elements
(TMEs) are controlled via surface gates using the alternating coupling
scheme with counter-intuitive pulse ordering (described below), a
variation of the CTAP protocol called Alternating CTAP (ACTAP).  This
geometry has been investigated in linear chains for quantum dots
\cite{PetrosyanOptComm2006} and in the context of a five donor Si:P
device \cite{JongNano2009}. Again, we do not wish to explore the
microscopic details of particular implementations too strongly,
instead focussing on the physics that can be demonstrated in all
potential platforms. Because of the wide range of possible
implementations, we do not treat decoherence here, although its effect
will be to wash out the interference patterns.

\section{Analytics and Modelling}
\begin{figure}[tb]
\includegraphics[width=0.45\textwidth]{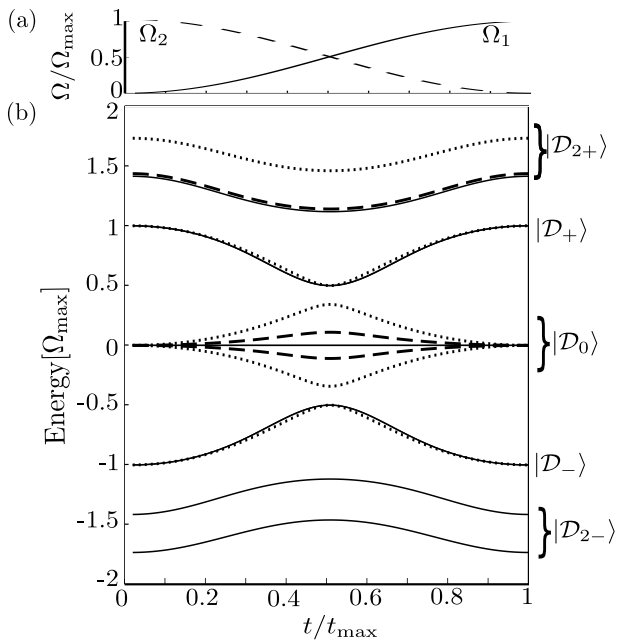}
\caption{(a) Tunnelling matrix elements $\Omega_1(t)$ and $\Omega_2(t)$, effecting the counter-intuitive pulse sequence. (b) Eigenspectra with varying
  detuning, $\Delta_u=\Delta_d=0$ (solid lines),
  $\Delta_u=-\Delta_d=0.25\Omega_{\max}$ (dashed) and $\Delta_u=-\Delta_d=\Omega_{\max}$ (dotted). The eigenstates can be ordered as indicated.  The double degeneracy of the null space is lifted with nonzero detunings of the central sites, we arbitrarily label the more energetic eigenstate of the pair $\ket{\mathcal{D}_0^{(+)}}$, and the less energetic $\ket{\mathcal{D}_0^{(-)}}$.
\label{fig:eigenspectra}}
\end{figure}

To explore the dynamics of the ACTAP interferometer, we write the Hamiltonian for this system as:
\begin{eqnarray}
\mathcal{H} & = & \Omega_1\left(\ket{1}\bra{2} + \ket{3_u}\bra{4} +
\ket{3_d}\bra{4}\right)  + h.c. \nonumber\\
& & \Omega_2\left(\ket{2}\bra{3_u} + \ket{2}\bra{3_d} +
 \ket{4}\bra{5}\right) +h.c. \nonumber \\
& & + \Delta_u\ket{3_u}\bra{3_u} + \Delta_d\ket{3_d}\bra{3_d}
\label{eq:hamiltonian}
\end{eqnarray}
where $\Omega_1$ and $\Omega_2$ are the gate controlled tunnelling
matrix elements (TMEs) and $\Delta_u$, $\Delta_d$ the energy detunings
of sites $\ket{3_u}$ and $\ket{3_d}$ respectively and $\hbar$ has been
set to 1. In general the TMEs between nearest neighbours will be independent,
  however by construction, we have chosen gate values that ensure the
  form of the Hamiltonian shown in Eq. 1.  This form is necessary to enforce the
  ideal ACTAP coupling scheme.  It should be realized, however, that
  complete symmetry is not required.  The ACTAP scheme has some
  robustness to variations from the ideal case, and this was explored
  for the linear ACTAP chain in Ref. \cite{JongNano2009}.  The choice of equality
  greatly simplifies the analysis and encapsulates all of the physics
  of CTAP.  Indeed this is the simplest coupling scheme that provides
  interferometry with a non-trivial loop and CTAP coupling. 
For simplicity, we choose squared sinusoid functions for the form of the
TMEs:

\begin{eqnarray}
\Omega_1 = \Omega_{1\max}\sin^2\left(\frac{\pi t}{2 t_{\max}}\right), \nonumber \\
\Omega_2 = \Omega_{2\max}\cos^2\left(\frac{\pi t}{2 t_{\max}}\right),
\end{eqnarray}
where $\Omega_{1\max}=\Omega_{2\max}=\Omega_{\max}$ is depicted in Fig.~\ref{fig:eigenspectra}(a). 

In the case $\Delta_u=\Delta_d=0$  the time dependent eigenvalues of
the Hamiltonian are 
\begin{eqnarray}
E_0 & = & 0,\nonumber\\
E_\pm & = & \pm \frac{\sqrt{3\Omega_1^2+3\Omega_2^2-\sqrt{\Omega_1^4+14\Omega_1^2\Omega_2^2+\Omega_2^4}}}{\sqrt{2}},\nonumber\\
E_2\pm & = & \pm \frac{\sqrt{3\Omega_1^2+3\Omega_2^2 + \sqrt{\Omega_1^4+14\Omega_1^2\Omega_2^2+\Omega_2^4}}}{\sqrt{2}},
\end{eqnarray}
where $E_0$ is doubly degenerate. These eigenvalues are
plotted in Fig.~\ref{fig:eigenspectra}(b), along with eigenspectra with non-zero detuning for
selected values of $\Delta_u=-\Delta_d$, for comparison. The
corresponding states are labelled $\ket{\mathcal{D}_0^{(+)}}$,
$\ket{\mathcal{D}_0^{(-)}}$, $\ket{\mathcal{D}_{\pm}}$, and
$\ket{\mathcal{D}_{2\pm}}$.  They are in general difficult to
represent in closed form, however the eigenvalues order naturally, and
these are indicated in Fig.~\ref{fig:eigenspectra}(b). In certain
limits we may extract some useful information about these states, and
some of these will be given below.  These plots illustrate the lifting
of the degeneracy of the $\ket{\mathcal{D}_0}$ states with detuning,
in these cases with opposite detunings, ie $\Delta_u=-\Delta_d$.

To enact the CTAP protocol we initialise the device so that the particle occupies site $\ket{1}$ at $t=0$, and apply an alternating pulse sequence in the
counter-intuitive ordering along the chain to effect
population transfer.   Conventional CTAP in linear devices (e.g. the straddling \cite{GreentreePRB2004,HollenbergPRB2006} or alternating schemes \cite{PetrosyanOptComm2006,JongNano2009}) works by adiabatically transforming the null state from this initial state to the desired output state.  Although in general there is no null state in our scheme, the adiabatic transfer still works analogously.
The evolution of this state under the counter-intuitive pulse sequence
gives rise to a smooth change in the population of the sites from
$\ket{1}$ at $t=0$ to $\ket{5}$ at $t=t_{\max}$. Since we are using
the alternating coupling scheme there is a transient occupation of
$\ket{3_u}$ and $\ket{3_d}$ during the protocol. In the straddling scheme
\cite{GreentreePRB2004,HollenbergPRB2006} occupation of all the
intermediate sites along the chain is strongly suppressed limiting its effectiveness for interferometric sensing.  Note that
as the degeneracy of the central sites, $\ket{3_u}$ and $\ket{3_d}$,
is broken by the gate induced detunings, interference will be observed
in certain regions of phase space. This interference is the effect that we are seeking to understand and exploit in this work.

\begin{figure}[tb]
\includegraphics[width=0.45\textwidth]{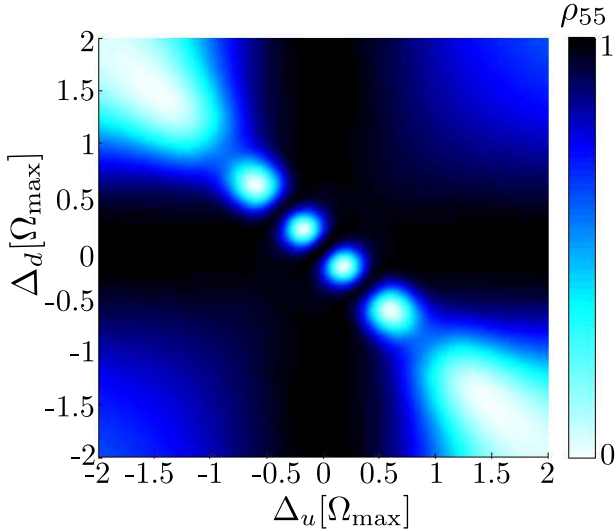}
\caption{(Colour online) Map of the final state population ($\rho_{55}$) for finite time showing
  the regions of adiabatic and non-adiabatic evolution. Dark regions indicate high fidelity transfer, the result of adiabatic evolution. Light regions indicate where the transport is low fidelity transfer. Note the overall `cross' of high-fidelity adiabatic transfer, with the EAB like oscillations along the line $\Delta_u = -\Delta_d$.
\label{fig:ctapinthires}}
\end{figure}

To examine the effect of the detunings on the final state population for the six site system
we numerically solve the Schr\"{o}dinger equation in Eq.~(\ref{eq:hamiltonian}) for the final state population of $\ket{5}$
after the CTAP pulse sequence has been performed in a finite time,
long enough to ensure adiabatic transfer in the zero detuning case,
as a function of the detunings on the middle sites.  The results of
this calculation are shown in Fig.~\ref{fig:ctapinthires}. This map of the final state population ($\rho_{55}$) shows
  several regimes of interest.  There is a prominent `cross' of
  high-fidelity transport where either $\Delta_u=0$ or $\Delta_d=0$.
  This region is governed by adiabatic timescales, and is discussed in
  Section III.  Superimposed on this, is a line of alternating high-
  and low-fidelity regions where $\Delta_u = -\Delta_d$.  This line
  corresponds to the EAB effect and is governed by nonadiabatic
  oscillations, and is discussed in Section \ref{sec:sixsite}.
\section{One arm with large detuning, other quasi resonant}\label{sec:onearm}
The main adiabatic feature of the system mapped out in Fig.~\ref{fig:ctapinthires} is the central cross region of high fidelity transfer. It is most clearly observed where at least one of the
middle site detunings is zero we have an adiabatic pathway that is similar to that of the simpler case with only one path as discussed in \cite{PetrosyanOptComm2006,JongNano2009}. To examine the width of the high fidelity region we consider the case where one detuning is taken to infinity as in this case we can retrieve some relatively simple analytical expressions. In this limit we recover a five site system with detuning on the middle site, and we explore the dependence of the adiabaticity on the detuning of the central site. Since the system is
symmetric, the width of the central cross feature is the same for the
vertical and horizontal axes. We are not aware of this case being treated previously in studies of alternating coupling schemes for adiabatic passage.  It is more conventional to consider detunings for states without any population (in our case the even numbered sites), or the end of chains, and so this case represents an interesting alternative detuning regime, which appears well suited to interferometric applications.

Without detuning the eigenstates are eigenvectors of the five site
chain can be fairly easily represented, as is done in
Ref.~\cite{JongNano2009}, but the central site detuning makes this
problematic. However, we can write down the states in appropriate limits, namely at the extrema of the protocol, and in the middle (when the adiabaticity is highest) as a series in the detuning, $\Delta$.  Note that for consistency, we shall keep the notation $\ket{\mathcal{D}_0}$ for the CTAP transport state, and the other eigenstates similarly labelled, however as mentioned above, there is no longer a null state with $\Delta \neq 0$.

\begin{widetext}

At the beginning of the protocol, $t=0$, we have $\Omega_1 = 0$ and the eigenvalues and eigenvectors are:
\begin{eqnarray}
E_0 &=& 0, \quad E_{\pm} = \pm{\Omega_2}, \quad E_{2\pm} = \pm \Omega_2 - \frac{\Delta}{2} + \mathcal{O}[\Delta]^2, \\
\ket{\mathcal{D}_0} &=& \ket{1}, \\
\ket{\mathcal{D}_{\pm}} &=& \frac{1}{\sqrt{2}}\left(\ket{4} \pm \ket{5}\right), \\
\ket{\mathcal{D}_{2\pm}} &=& \frac{\left(\pm 1 - \frac{\Delta}{2\Omega_2} + \mathcal{O}[\Delta]^2\right)\ket{2} + \ket{3}}{\sqrt{2 \pm \frac{\Delta}{\Omega_2} + \mathcal{O}[\Delta]^2}}, \\
\end{eqnarray}
and at the end of the protocol, $t=t_{\max}$, we have $\Omega_2 = 0$ and
\begin{eqnarray}
E_0 &=& 0, \quad E_{\pm} = \pm{\Omega_1}, \quad E_{2\pm} = \pm \Omega_1 - \frac{\Delta}{2} + \mathcal{O}[\Delta]^2, \\
\ket{\mathcal{D}_0} &=& \ket{5}, \\
\ket{\mathcal{D}_{\pm}} &=& \frac{1}{\sqrt{2}}\left(\ket{1} \pm \ket{2}\right), \\
\ket{\mathcal{D}_{2\pm}} &=& \frac{\left(\pm 1 - \frac{\Delta}{2\Omega_1} + \mathcal{O}[\Delta]^2\right)\ket{3} + \ket{4}}{\sqrt{2 \pm \frac{\Delta}{\Omega_1} + \mathcal{O}[\Delta]^2}}. \\
\end{eqnarray}
The more interesting case is at the midpoint of the transport protocol, i.e. $t=t_{\max}/2$, and for simplicity, setting $\Omega_1 = \Omega_2 = \Omega_{\max}/2$, we have
\begin{eqnarray}
E_0 &=& \frac{\Delta}{3} + \mathcal{O}[\Delta]^2, \quad E_{\pm} = \pm\frac{\Omega_{\max}}{2}, \quad E_{2\pm} = \pm \frac{\sqrt{3}\Omega_{\max}}{2} + \frac{\Delta}{3} + \mathcal{O}[\Delta]^2, \\
\ket{\mathcal{D}_0} &=& \frac{\left(\ket{1} - \ket{3} + \ket{5}\right) + \left(\frac{2\Delta}{\Omega_{\max}} + \mathcal{O}[\Delta]^2\right)\left(\ket{2} + \ket{4}\right)}{\sqrt{3}}, \\
\ket{\mathcal{D}_{\pm}} &=& \frac{1}{2}\left(\ket{1} \pm \ket{2} \mp \ket{4} - \ket{5}\right), \\
\ket{\mathcal{D}_{2\pm}} &=& \frac{\left(\ket{1} + \ket{5}\right) + \left(\pm \sqrt{3} + \frac{2\Delta}{3\Omega_{\max}} + \mathcal{O}[\Delta]^2\right)\left(\ket{2} + \ket{4}\right) + \left(2 \pm \frac{4\Delta}{\sqrt{3}\Omega_{\max}} + \mathcal{O}[\Delta]^2\right)\ket{3}}{\sqrt{12 + \frac{8 \Delta \sqrt{3}}{\Omega_{\max}} + \mathcal{O}[\Delta]^2}}. \\
\end{eqnarray}

\end{widetext}

\begin{figure}[tb]
\includegraphics[width=0.45 \textwidth]{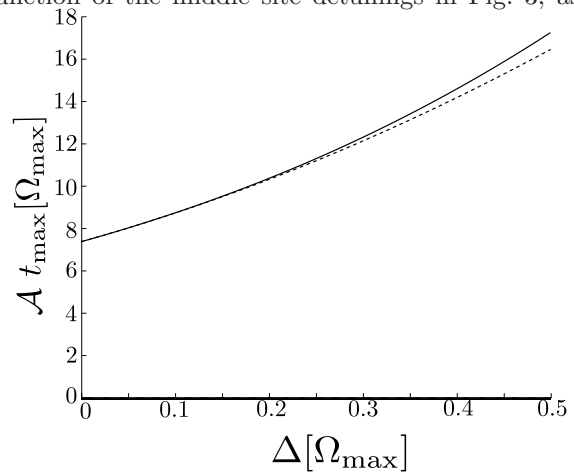}
\caption{Maximum adiabaticity through the protocol as a function of
  middle site detuning for A-CTAP in a 5 site system, corresponding to
  the case in the interferometer where one detuning has been taken to $\infty$.  The solid line is the numerical determination, Eq.~(\ref{eq:adiabaticity}) and the dashed is the analytical result to second order in $\Delta$, Eq.~(\ref{eq:adiabseries}).  The smooth increase in adiabaticity correlates to the smooth reduction in fidelity seen in Fig.~\ref{fig:ctapinthires} in the zones where $\Delta_u$ and $\Delta_d$ have the same sign.
\label{fig:deltavadiabatnum}}
\end{figure}

To quantify the adiabatic transfer in this five-site limit, and thereby gain insight over the adiabatic cross, we use the adiabaticity parameter defined for the five-site configuration
\begin{equation}
\mathcal{A}= \frac{|\bra{D_+}\frac{\partial\mathcal{H}}{\partial t}
  \ket{D_0}| }{\left| E_+-E_0 \right|^2}.
\label{eq:adiabaticity}
\end{equation}
For adiabatic evolution of the system we require $\mathcal{A} \ll 1$. Applying a detuning to the central site the adiabaticity changes and may shift the system out of the adiabatic regime. Numerically one sees the increase in the adiabaticity parameter with
increasing $\Delta$ as is shown in Fig.~\ref{fig:deltavadiabatnum}. High fidelity transfer (low $\mathcal{A}$)
occurs over a range of detunings. The effect of the detuning has
been examined for  STIRAP  \cite{VitanovARPC2001} and for straddling CTAP
in Ref. \cite{GreentreePRB2004}.  However it is worth noting that these cases are different from the one we are considering at present, as they other treatments examine energy shifts of states with vanishing occupation.  Here, the populations are explicitly non-zero due to the construction of the alternating coupling scheme, increasing the effect of the detuning.  This is an important feature and in fact \textit{necessary} to achieve an interferometric readout signature.
With the sinusoidal pulse scheme, the maximal value of the
adiabaticity occurs at $t=t_{\max}/2$, $\Omega_{1\max}=\Omega_{2\max}= \Omega_{\max}$ and $\Delta = 0$, we find \cite{JongNano2009}
\begin{equation}
\mathcal{A}=\frac{4 \pi}{\sqrt{3}\Omega_{\max}t_{\max}}.
\end{equation}

To include the effect of the detuning on the adiabaticity we perform a series expansion on the adiabaticity in $\Delta$. The adiabaticity parameter, to second order in $\Delta$ is  
\begin{eqnarray}
\mathcal{A}=\frac{4 \pi}{\sqrt{3} \Omega_{\max} t_{\max}} + \frac{20
  \pi \Delta}{3\sqrt{3} \Omega_{\max}^2 t_{\max}}\\ \nonumber
 + \frac{56 \pi \Delta^2}{9 \sqrt{3} \Omega_{\max}^3 t_{\max}}.
\label{eq:adiabseries}
\end{eqnarray}
This form of the adiabaticity parameter is plotted in Fig.~\ref{fig:deltavadiabatnum} and compared with a full numerical calculation of $\mathcal{A}$. 

\section{Interferometry in the six site system}\label{sec:sixsite}
\begin{figure}[tb]
\includegraphics[width=0.45\textwidth]{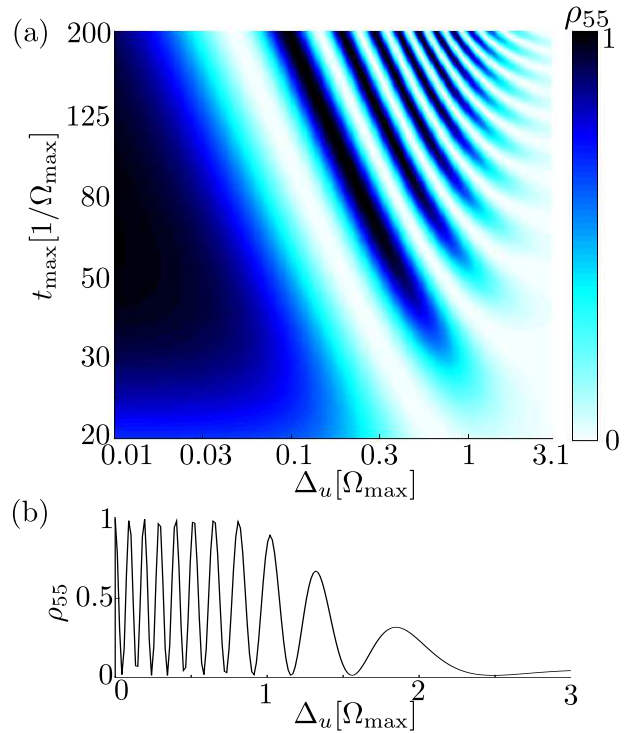}
\caption{(Colour online) (a) Final state population as a function of total time for anti-symmetric detunings ($\Delta_u=-\Delta_d$). As the total time increases we may resolve more instances of interference when the accumulated phase difference between the two paths is $\pi /2$. (b) Trace of the final state population as a function of $\Delta_u$ for the maximum time shown in (a).
\label{fig:tmaxvdelta}}
\end{figure}

Returning to the map of final state population as a function of the
middle site detunings in Fig.~\ref{fig:ctapinthires}, as well as the
central cross feature there is also the line of low-fidelity regions across the $\Delta_u=-\Delta_d$ diagonal where we see non-adiabatic evolution. Along this line we see regions of low fidelity
transfer superimposed upon the region where we expect high
fidelity. The position and number of these areas vary with the total
time, as is shown in Fig.~\ref{fig:tmaxvdelta}. 
We see that as the total time is increased, the frequency of
the nonadiabatic oscillations increases with increasing detuning. 
\begin{widetext}
We can understand the interference by considering the eigenspectrum with non-zero detuning, in particular focusing on the lifting of the null state degeneracy.  As we see in Fig. \ref{fig:eigenspectra}(b) when the sites are oppositely detuned
  ($\Delta_u = -\Delta_d$), the states $\ket{\mathcal{D}_0^{(+)}}=\ket{\mathcal{D}_0^{(-)}}$ are degenerate at the start and end, but
  are split evenly during the protocol and it is the population
  oscillations between these states which will be important, rather
  than the adiabaticity treated in Eq.~(\ref{eq:adiabseries}). To first order in $\Delta$, at the midpoint of the protocol we find
\begin{eqnarray}
E_0^{(\pm)} &=& \pm \frac{\Delta}{\sqrt{5}}, \\
\ket{\mathcal{D}_0^{(\pm)}} &=& \frac{1}{\sqrt{5}}\left[\left(\ket{1} + \ket{5}\right) \pm \frac{\Delta}{\sqrt{5}\Omega_{\max}}\left(\ket{2} + \ket{4}\right) + \frac{1}{2}\left(-1 \mp \sqrt{5}\right) \ket{3_u} + \frac{1}{2}\left(-1 \pm \sqrt{5}\right)\ket{3_d}\right].
\end{eqnarray} 
\end{widetext}
 
The interference observed in the final state population can be understood as resulting from the electron undergoing Landau-Zener oscillations between these two states as the Hamiltonian is evolving with the protocol. The non-adiabatic behaviour can be thought of as arising due to the total protocol time required for the system to be able to resolve these two states. That is, when the total time is large compared to the energy gap between them the system is able to resolve the lifted degeneracy of the null states. This energy gap depends on the detunings of the central sites. Empirically, we find for a given time $t_{\max}$, maxima in the transfer fidelity occur when
\begin{equation}
\Delta_n=  \frac{f n}{ t_{\max}}
\end{equation} 
where $\Delta_n$ is the detuning of the $n^{\mathrm{th}}$ maximum with
$\Delta_n = \pm \Delta_u = \mp \Delta_d$ and a factor $f\sim20$ which
has been determined empirically.
Since the states we are interested in are approximately parallel, a doubling in the total time will correspond approximately to a doubling in the frequency. Since these oscillations are periodic the non-adiabatic behaviour will also be periodic in the total time. This is plotted in Fig.~\ref{fig:tmaxvdelta} where we see very clear linear relations on a log-log scale, except in the large $\Delta$ limit.  With increasing detuning there is an overall reduction in
  fidelity, as the barrier between sites 1 and 5 increases, thereby
  suppressing population transfer. This effect is most evident in the lower fidelity transport regions outside the central cross feature in Fig.~\ref{fig:ctapinthires}.
\begin{figure}[tb]
\includegraphics[width=0.45\textwidth]{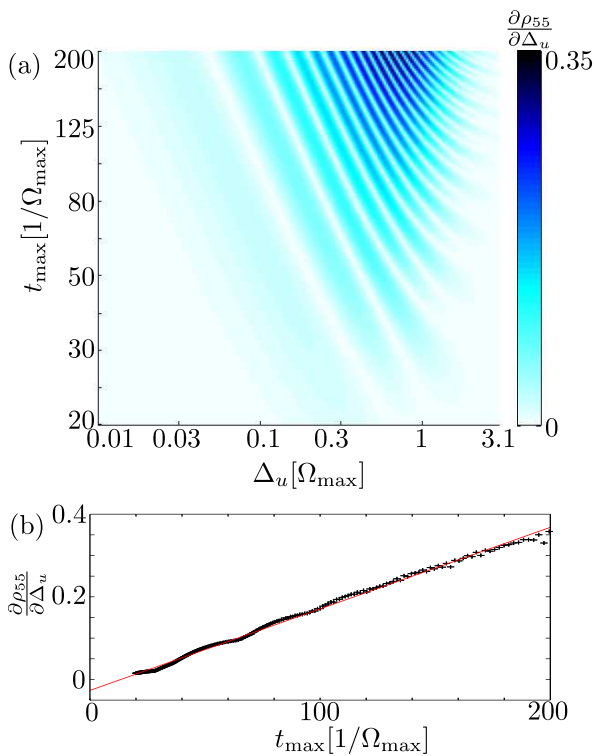}
\caption{(Colour Online)(a)Derivative of the final state population with respect to $\Delta_u$. Regions with a large rate of change may be considered useful for charge sensing applications. The very light regions are of high negative derivative and dark regions high positive derivative. The presence of a charge will alter the final state population due to its effect on the energy of the detuned site $\ket{3_u}$ or $\ket{3_d}$ in the arm being used as the sensor. (b) Dependence of the sensitivity (derivative of the final state population) on $t_{\max}$.\label{fig:sensitivity}}
\end{figure}

The distinct regions of adiabatic and non-adiabatic behaviour suggest that such a device might be interesting for charge sensing. Mapping out the rate of change of the final state population with respect to one detuning, as shown in Fig.~\ref{fig:sensitivity}, we can identify device detuning configurations which will display a large response to any change in the energy level of the one of the middle sites due to a charge being in close proximity.  Using one arm as the detector arm and selecting a detuning configuration which will show this large response we can see that the final state populations respond to the presence of a near-by charge.  Fig.~\ref{fig:sensitivity}(b) also shows the dependence of the sensitivity on time for the first fringe from Fig.~\ref{fig:sensitivity}(a). The sensitivity of the interferometer at these operating points scales linearly with the total protocol time.
\section{Conclusion}
Mapping of the final state population of a 6 site CTAP interferometer in an electrostatic Aharonov-Bohm style geometry shows an interesting interplay between adiabatic and non-adiabatic transport of an electron in the device. We may tune the device between these two regimes easily by modifying the total protocol time for the CTAP transport or by changing the on-site energy of the middle sites in the two different paths. We have modelled the behaviour of this device and the dependence of the detuning on the adiabaticity. We see effects similar to electrostatic Aharonov-Bohm oscillations as seen in the periodic behaviour of the final-state population when the two paths have opposite detuning. In some configurations a small alteration to the energy of a site will result in the device being shifted from adiabatic transfer to non-adiabatic. This is seen in a large change in the final state population. Such sensitivity suggests a device which may also offer opportunities for charge sensing applications.
\begin{acknowledgments}
We thank Andy Martin and David Jamieson for useful discussions and careful proof reading of the manuscript.  Andrew Greentree acknowledges the Australian Research Council for financial support under project no. DP0880466.
This work was supported by the Australian
Research Council, the Australian Government, and the US National Security Agency
(NSA) and the Army Research Office (ARO) under contract number W911NF-08-1-0527.
\end{acknowledgments}

\bibliography{Jong_interferometer}

\end{document}